# Particle Ratios in a Multi Component Non-Ideal Hadron Resonance Gas

**Rameez Ahmad Parra[a], Saeed Uddin[b], Waseem Bashir[c] and Inam-ul Bashir[d]**

[a] Department of Physics, Central University of Kashmir, India

[b] Department of Physics, Jamia Millia Islamia, New Delhi, India

[c] Department of Physics, Govt. Degree College, Budgam, J&K, India

[d] Govt. Boys Higher Secondary School, Tral, J&K, India

## Abstract

We have considered formation of a multi component non-ideal hot and dense gas of hadronic resonances in the ultra-relativistic heavy-ion collisions. In the statistical thermal model approach the equation of state (EoS) of the non interacting ideal hadron resonance gas (IHRG) does not incorporate either the attractive part or the short range repulsive part of the baryonic interaction. On the other hand in the non-ideal hadron resonance gas (NIHRG) model we can incorporate these interactions using the Van der Waals (VDW) type approach. Studies have been made to see its effect on the critical parameters of the quark-hadron phase transition. However, it can also lead to modifications in the calculated relative particle yields. In this paper we have attempted to understand the effect of such Van der Waals type interactions on the relative particle yields and also studied their dependences on the system's thermal parameters, such as the temperature and baryon chemical potential ($\mu_B$). We have also taken into account the decay contributions of the heavier resonances. These results on particle ratios are compared with the corresponding results obtained from the point-like i.e. non interacting IHRG model. It is found that the particle ratios get modified by incorporating the Van der Waals type interactions, especially in a baryon rich system which is expected to be formed at lower RHIC energies, SPS energies and in the forthcoming CBM experiments due to high degree of nuclear stopping in these experiments.

## Introduction:

The particle production in the hot and dense hadronic medium created in the ultra-relativistic heavy ion collisions enables us to understand the final stage properties of such a hadronic matter. The matter which existed at the very early stages of universe can be recreated and studied in these collisions, though at a very small length scale. It is believed that the matter created in such nuclear collisions in the laboratory can achieve reasonably high degrees of thermal and chemical equilibrium. The study of the hadronic yields in the heavy-ion collisions at ultra-relativistic energies is therefore an important tool to explore the properties of the hadronic matter, also called fireball, formed in these collisions [1-9].

The statistical thermal models provide a suitable framework to study the final stage properties of the hot fireball formed in the ultra-relativistic nucleus-nucleus collisions [10-18]. The hot and dense secondary partonic matter (consisting of quarks and gluons) initially produced within the fireball is in a pre-equilibrium state. These secondary partons continue to undergo multiple elastic and inelastic collisions. The fireball initially formed in the collision thus grows in size due to multi-particle production. It also simultaneously develops a hydrodynamic expansion due to the collective effect. The system thus reaches a state of sufficiently high degree of thermal and chemical equilibrium with sufficiently large number of particles within it which then permits the application of the statistical thermal models. If the initial temperature (T) and net baryon density ($\rho$) is sufficiently high i.e. $T > T_C$ (or $\rho > \rho_c$) then this may result in the formation of quark gluon plasma (QGP) phase [19-22]. This state is then followed by a mixed phase of QGP + HRG, assuming a first order phase transition. After the completion of a first order phase transition process, the particles (i.e. the hadrons) in the HRG phase still continue to interact. The collisions among the produced hadrons can lead to a high degree of thermal and chemical equilibrium of various hadronic species in the HRG phase as well. The HRG system continues to grow, expand and cool and in the process gets diluted. After this a freeze-out occurs when the mean free paths of these particles become comparable with the overall size of the system [23-26].

It has been highlighted earlier that the point-like hadrons do not reproduce the ground state properties of the nuclear matter. Further no reasonable first order QGP – HRG phase transition within the framework of a thermal model can be constructed with suitably large number of degrees of freedom in the HRG phase [27-30]. This essentially happens because a large number of point-like hadronic resonances can be thermally excited at high temperatures in absence of any repulsive interaction in a given physical volume of the system. Consequently, at adequately high temperature the pressure in the ideal HRG phase pressure becomes more than the pressure in the QGP phase (i.e. $P_{HRG}$ > $P_{QGP}$). Or in simple words, considering a first order quark-hadron phase transition in a statistical thermal model framework with sufficiently large number of hadronic degrees of freedom, the system at high temperatures reverts to the hadronic resonance gas (HRG) phase [31-35]. According to lattice quantum chromodynamics (LQCD) predictions for vanishing net baryon number, the phase transition occurs at the critical temperature around 160 MeV and the system remains in the QGP phase at further higher temperatures [36-38]. Therefore, the result obtained by considering the HRG system consisting of point-like hadrons contradicts the LQCD prediction. However, within the framework of the statistical thermal models, this problem can be solved by considering the repulsive interactions between baryons (or antibaryons), which give rise to an excluded volume type effect. Moreover, in some phenomenological models the attractive and repulsive interactions both have been taken into account where the strength of the repulsive interaction is proportional to the net baryon number density ($n$). Consequently, it this interaction ceases to exist for $n \cong 0$ [39-41]. We have employed grand canonical ensemble (GCE) partition function approach to describe the properties of a gas of interacting hadronic resonances. The canonical ensemble partition function approach doesn't give broad thermodynamical picture of the hadronic system, because the number of particles and total energy are usually not conserved in the real physical systems formed in the ultra-relativistic nuclear collisions. The interactions in the multi-component hadron resonance gas (HRG) can be incorporated in a more appropriate way by using the grand canonical ensemble (GCE) partition function approach. The repulsive part of interaction has been earlier incorporated through an excluded volume type effect in the HRG model and is generally known as EV-HRG model of heavy ion collisions [42-44]. On the other hand, the attractive part of interaction is incorporated by way of multiplying the grand partition function with an exponential factor containing an average attractive potential. The attractive and repulsive interactions can be incorporated by other ways also, such as in the Non-linear Walecka model and its generalizations [45-49]. In case of the classical Van der Waals type approach, the strength of repulsion is proportional to the total particle density and is therefore finite even at zero baryon chemical potential ($\mu_B$

= 0). Conversely, in the case of Walecka model, the strength of repulsion is proportional to the *net* baryon density and hence it vanishes at $\mu_B$ = 0. This is therefore a significant advantage of Van der Waals type approach over the Walecka approach. In the following we will review this aspect to arrive at the modified equation of state (EoS) of such a non-ideal hadron resonance gas (NIHRG) system.

## 2. The Statistical Approach

In this section we will attempt to provide an overview of the Van der Waals-type EoS for interacting NIHRG which is done by employing the grand canonical ensemble (GCE) formulation with both attractive as well as repulsive interactions taken into account. The repulsive force is assumed to exist between pairs of two baryons and pairs of two antibaryons, while it is purely attractive between a baryon-antibaryon pair. The ideal (i.e. interaction free) grand canonical partition function is given by:

$$\mathcal{Z}(T,\mu,V) = \sum_{N=0}^{\infty} e^{\frac{\mu N}{T}}\, Z(T,N,V) \tag{1}$$

Here $Z(T,N,V)$ represents the canonical partition function for $N$ number of particles. T and $V$ are the temperature and total physical volume of the system, respectively. The phenomenological grand partition function with both the attractive and repulsive interactions taken into account can be written as:

$$\mathcal{Z}^{int}(T,\mu,V) = \sum_{N=0}^{\infty} e^{\frac{\mu N}{T}}\, Z(T,N,V-bN)e^{-\frac{\overline{U}}{T}} \tag{2}$$

Here $N$ is the number of particles in the system with both attractive and repulsive interactions present (i.e. $N = N^{int}$). Here $bN$ represents the excluded volume of baryons (antibaryons) arising due to hard-core repulsive interaction [27,31-33]. The attractive interaction is taken into account in the form of an average interaction energy through the introduction of the factor $e^{-\overline{U}/T}$ in the grand partition function, where the $\overline{U}$ represents the average attractive interaction energy assuming a uniform particle density, $n = N/V$. The mean attractive interaction energy can be written as $\overline{U} = \frac{1}{2}\sum_{i,j} V_{att}(\vec{r}_i - \vec{r}_j)$, where $V_{att}(\vec{r}_i - \vec{r}_j)$ represents the average attractive interaction energy of any two particle pair, which depends on their relative coordinates $(\vec{r} = \vec{r}_i - \vec{r}_j)$. Contribution of three or more particle interactions is ignored because such collisions would be rare. Thus in general it is sufficient to consider only one pair of particles [39]. Under the assumption of a uniform particle density $n = N/V$ the total interaction energy can be obtained by summing over all the particles pairs. Therefore (for large $n$) we can write:

$$\overline{U} = \frac{1}{2}\int d^3\vec{r}_1 d^3\vec{r}_2 n(\vec{r}_1)\, n(\vec{r}_2) V_{att}(\vec{r}_2 - \vec{r}_1) \quad (3)$$

Which finally yields $\overline{U} \equiv \frac{n^2 V u}{2}$. The quantity $u = \int d^3\vec{r}\, V_{att}(\vec{r})$. By defining $a = \int 2\pi r^2 dr\, V_{att}(\vec{r})$, where "$a$" represents the effective attractive parameter, we can write $u = 2a$. Using Equation (3) we finally get $\overline{U} = n^2 V a$

We can therefore rewrite:

$$\mathcal{Z}^{int}(T,\mu,V) = \sum_{N=0}^{\infty} e^{\frac{\mu N}{T}} \mathcal{Z}(T,N,V-bN) e^{-\frac{n^2 V a}{T}} \quad (4)$$

Where $\mathcal{Z}(T,N,V-bN)$ is the canonical partition function taking into account the excluded volume effect arising due to hard-core hadronic repulsion. The pressure function is obtained [27] by identifying the extreme right hand singularity of the Laplace transform of $\hat{\mathcal{Z}}^{int}$ in equation (4):

$$\hat{\mathcal{Z}}^{int}(T,\mu,\zeta) = \int e^{-\zeta V} \mathcal{Z}^{int}(T,\mu,V) dV \quad (5)$$

The $\varsigma$ is the parameter of Laplace transformation. One can rewrite for uniform particle number density:

$$\hat{\mathcal{Z}}^{int}(T,\mu,\zeta) = \int e^{-\zeta V}\ \mathcal{Z}^{excl}(T,\mu,V) e^{-\frac{n^2 V a}{T}} dV \quad (6)$$

Where $\mathcal{Z}^{excl}(T,\mu,V) = \sum_{N=0}^{\infty} e^{\frac{\mu N}{T}} \mathcal{Z}(T,N,V-bN)$ is the grand canonical partition function with excluded volume effect taken into account.

The extreme right hand singularity of the Laplace transformation in equation (6) can be located by rewriting the integrand in equation (6) to get:

$$\hat{\mathcal{Z}}^{int}(T,\mu,\zeta) = \int e^{-V\left(\zeta - \frac{ln\mathcal{Z}^{excl}(T,\mu,V)}{V} + n^2 a/T\right)} dV \quad (7)$$

The finiteness of the integral in equation (7) requires that in the limit $V \to \infty$ the extreme right hand singularity must satisfy:

$$\zeta = \frac{1}{T} \lim_{V\to\infty} \left[\frac{T ln\mathcal{Z}^{excl}(T,\mu,V)}{V} - n^2 a\right] \quad (8)$$

Defining $\qquad p^{excl}(T,\mu,V) = \lim_{V\to\infty} T\left\{\frac{\{ln\, \mathcal{Z}^{excl}(T,\mu,V)\}}{V}\right\}$ and $\zeta T = p^{int}$

We can then finally write the pressure of the system incorporating the attractive as well as repulsive interaction:

$$p^{int}(T,\mu,V) = p^{excl}(T,\mu,V) - an^2 \quad (9)$$

It may be noted that here $n = n^{int}$. By taking into account the effect of repulsive interaction, leading to an excluded volume type effect, we can write [27,33] by considering the available volume in the system

$$p^{excl}(T,\mu,V) \ = \frac{nT}{1-bn}$$

This gives $$p^{int}(T,\mu,V) = \frac{nT}{1-bn} - an^2 \quad (10)$$

The particle number density can then be obtained in a thermodynamically consistent way by using $n = \left.\partial p^{int}/\partial\mu\right|_T$. This finally yields a relation for the chemical potential:

$$\mu = Tln\left(\frac{n}{1-bn}\right) + \frac{T}{1-bn} - 2an + C \quad (11)$$

With $$C = \mu - Tln\left(n^{id}\right) - T$$

Writing the point-like particle number density $n^{id} = e^{\mu/T}\phi$, we get:

$$C = Tln\left(\frac{n^{id}}{\phi}\right) - Tln\left(n^{id}\right) - T \quad (12)$$

Thus we get $$\mu = Tln\left\{\frac{n}{(1-bn)\phi}\right\} + \frac{nTb}{1-bn} - 2an \quad (13)$$

Defining, $\mu^* = Tln\left\{\frac{n}{(1-bn)\phi}\right\}$ and writing $\phi$ in terms of $\mu^*$ i.e. $\phi = n^{id}(T,\mu^*)e^{-\mu^*/T}$ and using this in equation (13), we will get an expression for the "effective" chemical potential, i.e. $\mu^*$ and the particle number density $n$ ( i.e. $n^{int}$) as:

$$\mu^* = \mu - \frac{nbT}{1-bn} + 2an \quad (14)$$

$$n = (1-bn)n^{id}(T,\mu^*) \quad (15)$$

This finally yields particle number density in the NIHRG model as:

$$n(\mu,T) = \frac{n^{id}(T,\mu^*)}{1+bn^{id}(T,\mu^*)} \quad (16)$$

The above equation (16) is the number density for single component of hadronic matter. For multi-component baryonic matter, we can generalize the above equation to get:

$$n_j(\mu_j,T) = \frac{n_j^{id}\left(T,\mu_j^*\right)}{1+\sum_i b_i n_i^{id}\left(T,\mu_j^*\right)} \quad (17)$$

Equation (17) represents the modified number density of the $j^{th}$ hadronic specie for a multi component NIHRG, using Van der Waals type equation of state (EoS). The excluded volume arising due to the $j^{th}$ hadronic species considering it a hard sphere is $(b_j = \frac{16}{3}\pi r_j^3)$, where $r_j$ represents the radius of the $j^{th}$ hadronic species. The summation over the index $i$ in the denominator also includes $j$. The effect of the attractive and the repulsive hard-core interaction appears in the equation of state of the system through the "effective" baryon chemical potential $(\mu^*)$ in equation (14).

The application of equation (14) to the anti-baryonic sector requires that the "effective" chemical potential for antibaryon (i. e. $\overline{\mu^*}$) be written as:

$$\overline{\mu^*} = \bar{\mu} - \frac{\bar{n}bT}{1-b\bar{n}} + 2a\bar{n} \quad \text{with} \quad \bar{\mu} = -\mu$$

Where the quantities with bar indicate their values for the anti-baryons. For the anti-baryonic sector in the multi-component NIHRG we will get:

$$n_{\bar{j}} = \frac{n_{\bar{j}}^{id}(T,\overline{\mu_j^*})}{1+\sum_i b_{\bar{i}}\, n_{\bar{i}}^{id}(T,\overline{\mu_j^*})} \tag{20}$$

For a baryon free matter, $\bar{\mu_j} = -\mu_j = 0$. This also gives $\overline{\mu_j^*} = \mu_j^*$, thus providing $n_j = n_{\bar{j}}$ at any given temperature T, which is mandatory to maintain the baryon–antibaryon symmetry in the system when $\mu_j = 0$. In the following we present results of our numerical calculations and discussion.

## 3. Results and Discussion

Making use of the above results we have calculated relative particle yields which may be compared with the experimental data at various energies such as SPS, RHIC etc. in order to determine the final stage freeze-out conditions in the fireballs formed in these experiments.

With this aim we have applied the above formulation for a system of hot and dense hadronic matter consisting of several hadronic species. The repulsive forces are assumed to exist between pairs of two baryons (fermions) and pairs of two antibaryons (anti-fermions), while it is purely attractive between baryon-antibaryon pairs. This is consistent with earlier approaches [31-33]. We have calculated various particle ratios and studied their variations with changes in temperature (T) and baryon chemical potential $(\mu_B)$ of the system. This is essentially done to see the effect of the Van der Waals type interaction on the number densities of different hadronic species as well as their relative abundances for the non-ideal multi-component hadron resonance gas (NIHRG) and compare the above results with those which are obtained for the point-like ideal hadron resonance gas (IHRG). The results are found to depend on the system's temperature and chemical potential.

At the collision energy ranges of the experiments at RHIC, SPS and the also in upcoming CBM experiment, the nuclear stopping effect is expected to be fairly high leading to a baryon rich fireball. The baryon rich fireballs are expected to maintain large chemical potentials which will vary with the degree of stopping in the given experiment. Variation in system's temperature will also be observed depending on the collision scenario, including system size [25,26].

In many earlier works employing, different model approaches the values of the attractive parameter $a$ have been used in the range 329-1250 MeV-fm$^3$. The repulsive parameter $b = (\frac{16}{3}\pi r_0^3)$ was fixed by choosing a suitable value of the hadronic hard-core radius $r_0$. We have used a constant value of the attractive parameter $a$ = 329 Mev-fm$^3$ and $r_0 =$ 0.59 fm. These values of the parameters in the Van der Waals type EoS are fixed by the ground state properties of nuclear matter [49-54]. In another approach $r_0 =$ 0.7 fm has also been used in a meson mean-field type model EoS to determine the ground state nuclear matter incompressibility [27].

While attempting to describe the properties of the hadronic matter formed in the ultra-relativistic nucleus-nucleus collisions we are essentially dealing with a system of strongly interacting particles hence the quantities such as baryon number and strangeness are conserved within the system. The weakly decaying particles such as lambda, sigma, cascade etc. also contribute to the observed lighter hadron multiplicities. However, at the strong interaction time scale these decays take place long after the freeze-out of the strongly interacting hadronic matter has taken place Moreover, as already discussed in the previous section, the total number of particles within the system before its final break-up is not fixed due to continuous creation, annihilation and other reaction processes. Thus the conservation of mean baryon number and the mean net strangeness content of the system is achieved within the framework of the grand canonical ensemble. This is done by introducing the baryon and strange chemical potentials, $\mu_B$ and $\mu_s$, respectively. Considering three quark flavours, we have in our calculation defined the chemical potentials of a given $j^{th}$ hadron as [17,18,35,55-61]:

$$\mu_i = (q_i - \bar{q}_i)\, \mu_q + (s_i - \bar{s}_i)\, \mu_s = N_q \mu_q + N_s \mu_s$$

Where $N_q$ and $N_s$ are the number of valence light (u,d) and strange (s) quarks, respectively in a given $j^{th}$ type hadronic specie with $\mu_q = \mu_B/3$.

In the following we present our results for the dependence of the "effective" baryon chemical potential, $\mu_B^*$, and various "relative" hadronic yields. We discuss their dependence on the thermal parameters of the system i.e. T and $\mu_B$. For this purpose we have included in our system hadronic resonances up to the omega mass (1672 MeV) and used their known weak decay channels to calculate their contributions to the lower mass hadronic multiplicities after the thermo-chemical freeze-out of the system has occurred.

We have determined the effective baryon chemical potential $\mu_B^*$ by solving the transcedental equations (14) and (17). In figure 1 we have shown the variation of $\mu_B^*$ with temperature for a baryon rich system by choosing $\mu_B =$ 300 MeV and 500 MeV. We find that as the temperatures increases beyond 80 MeV, the $\mu_B^*$ shows a very slightly increase but beyond T $\sim$ 120 – 130 MeV it starts decreasing rapidly.

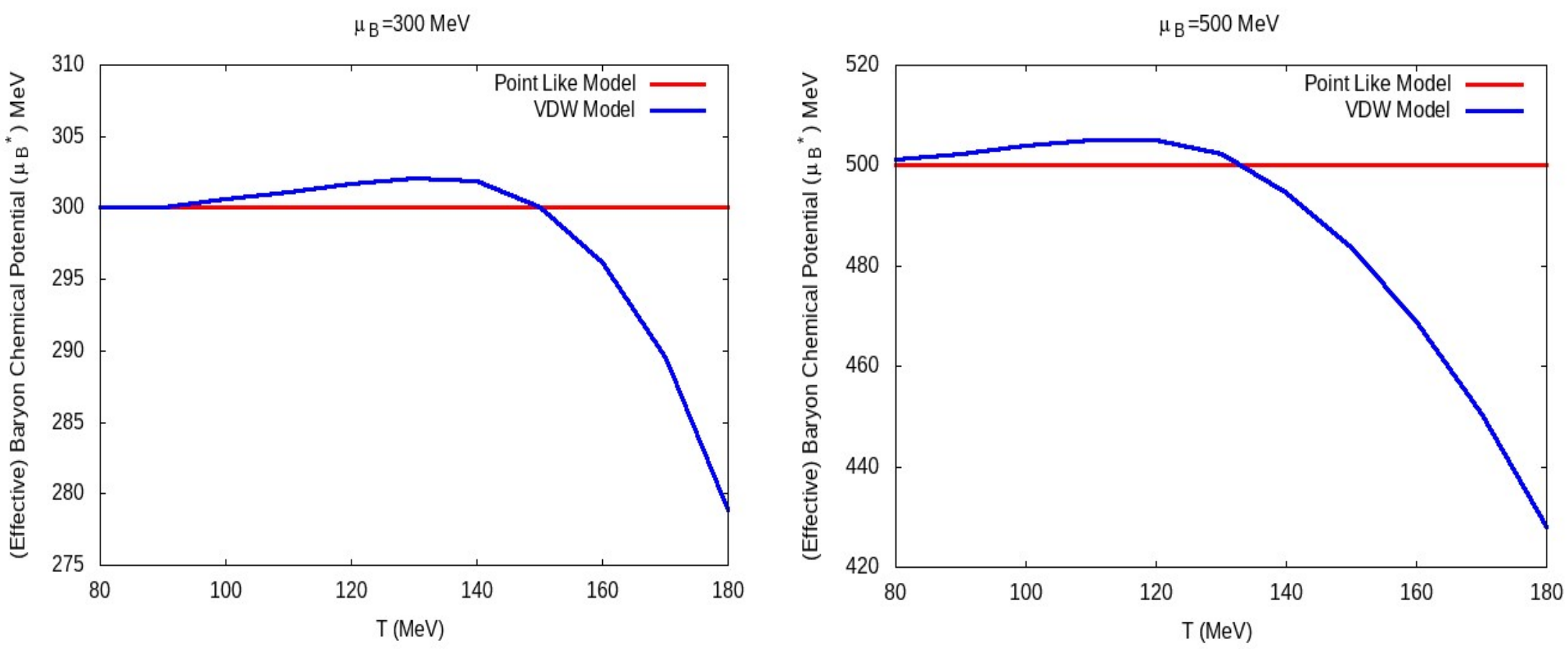

**Fig 1: Variation of "effective" baryon chemical potential with temperature, T, in a baryon rich system for two fixed values of baryon chemical potential, $\mu_B$ = 300 MeV and 500 MeV.**

This effect is further highlighted in figure 2 where we have shown the dependence of the "relative" effective baryon chemical potential i.e. $\mu_B^*/\mu_B$ with T for $\mu_B$ = 300 MeV and 500 MeV.

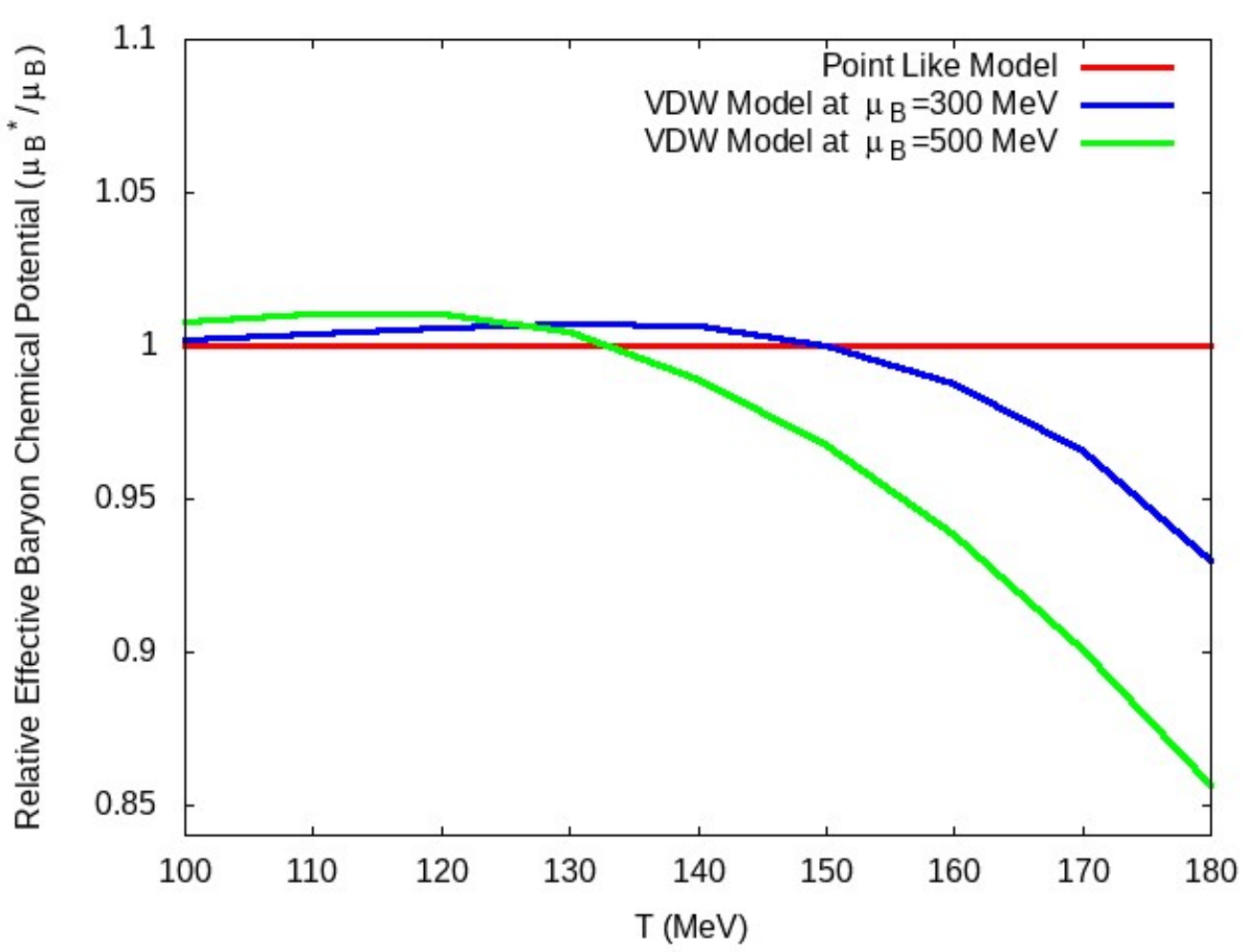

**Fig 2: Variation of “relative” effective baryon chemical potential with temperature, T, in a baryon rich system for two fixed values of baryon chemical potential, $\mu_B = 300$ MeV and 500 MeV.**

In the ultra-relativistic nucleus-nucleus collisions the antiproton to proton ratio ($\bar{p}/p$) is considered an important indicator of the degree of nuclear stopping. This also allows us to learn how baryon numbers initially carried by the nucleons only, before the nuclear collision, are distributed in the final state [62-68]. With this aim we have in figure 3 shown the dependence of anti-proton to proton ratio ($\bar{p}/p$) on temperature at two fixed values of baryon chemical potential, $\mu_B$ = 300 MeV and 500 MeV. For each case (viz. the point-like IHRG and NIHRG cases) we have considered two situations in our calculations i.e. when the weak decay contributions of heavier resonances after the freeze-out are included and when they are ignored. The $\bar{p}/p$ ratio increases with increasing temperature for all cases. This essentially happens since the number of $\bar{p}$ and the hadrons (mostly anti-baryons) decaying after the freeze-out contributing to the final state $\bar{p}$ multiplicity increases due to their enhanced thermal production, which was otherwise suppressed due to high chemical potential in a baryon rich system. The decay contributions are seen to further increase this ratio compared to the situation when they are not taken into account. In the case of the NIHRG the $\bar{p}/p$ ratio is seen to increase appreciably. Thus the interactions incorporated in the NIHRG system may play an important role in determining the actual freeze-out conditions in a hot baryon rich system i.e. at sufficiently large values of $\mu_B$ and T.

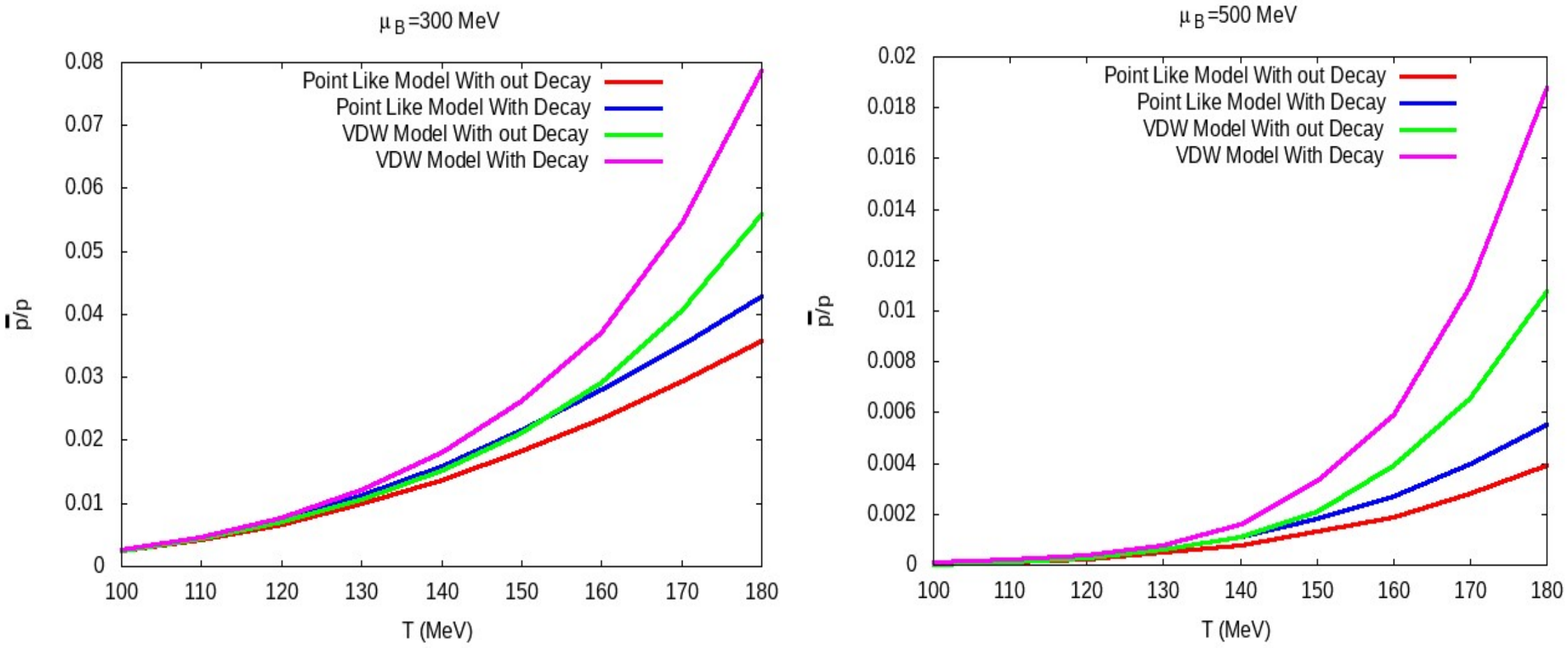

**Fig 3: Dependence of the $\overline{p}/p$ ratio on temperature for $\mu_B = 300$ MeV and 500 MeV.**

The mid-rapidity $\bar{p}/p$ ratio in central Pb + Pb collisions at 40A GeV in CERN-SPS experiments is ≈ 0.0078 [69,70]. In the above results if we chose $\mu_B$ = 500 MeV and T ~ 160 MeV obtain a good description of this ratio if we use the equation of state of NIHRG. Similarly the experimental mid-rapidity $\bar{p}/p$ ratio at 80A GeV is ≈ 0.028 [70] and can be obtained for $\mu_B$ = 300 MeV and T ~ 155. In all thermal models the value of the baryon chemical potential $\mu_B$ essentially indicates the excess of baryons over antibaryons. We therefore expect the system to maintain a larger chemical potential at 40A GeV as compared to 80A GeV due to relatively lesser thermal production of antibaryons (including antihyperons) hence leading to a relatively large excess of baryons over antibaryons at 40A GeV as compared to 80A GeV.

With non-consideration of the resonance decay and more importantly the interaction in the HRG, one would need a significantly large value of temperature at this value of chemical potential to reproduce the $\bar{p}/p$ ratio value which may seem somewhat unphysical. However, one may still get a reasonable value of this ratio in IHRG case at a lower value of T by choosing a smaller value of $\mu_B$.

As indicated at these energies since the nuclear stopping is effective hence the system formed in such collisions are expected to be baryon rich i.e. with high net baryon density and therefore will maintain sufficiently large chemical potential [23, 62, 71-77]. Hence the determination of the values of chemical potential and temperature of the system at freeze-out can help us understand the degree of nuclear stopping. As one moves up from SPS to RHIC energy, approximately the same temperature but a significantly smaller baryonic chemical potential is observed in the central rapidity region [23,62,78]. Theoretical extraction of these values will depend on the choice of the statistical model used for analyzing the experimental data.

Experiments have shown enhancement in the overall strangeness production in the nuclear collisions at ultra-relativistic energies relative to the non-strange hadrons. Therefore experimental measurement of the strange hyperons created in nuclear collisions is also an important tool to study the properties of the hot and dense systems typically produced in nuclear collisions at ultra-relativistic energies. Unlike hadron–hadron collisions, we expect that in the most central ultra-relativistic nuclear collisions the *participating* quarks will scatter many times before joining in an asymptotic hadronic state at the stage of hadronization. Application of the well-established methods of statistical physics can provide a simplified approach to theoretically predict the strangeness abundance [79]. Such attempts have been made earlier also to describe strangeness production in ultra-relativistic nucleus-nucleus collisions in terms of an equilibrated gas of hadronic resonances [57-61].

In the statistical thermal approach the abundances of strange particles in the system is affected not only by the values of $\mu_B$ and T but also by the value of the strange chemical potential, $\mu_S$, which, as already stated previously, is fixed by the strangeness conservation criteria for the given values of $\mu_B$ and T.

In figure 4 we have plotted the variation of "effective" *strange chemical potential* $(\mu_S^*)$ with temperature T for $\mu_B = 300$ MeV and 500 MeV. The strange baryon chemical potential value obtained using Van der Waals type interaction in NIHRG model is compared with the point-like hadron case (i.e. IHRG). We notice that after initial decrease the "effective" strange chemical potential $(\mu_S^*)$ gets enhanced in presence of Van der Waals (VDW) type interactions in the HRG as compared in point-like hadron case, especially at higher temperatures and chemical potentials.

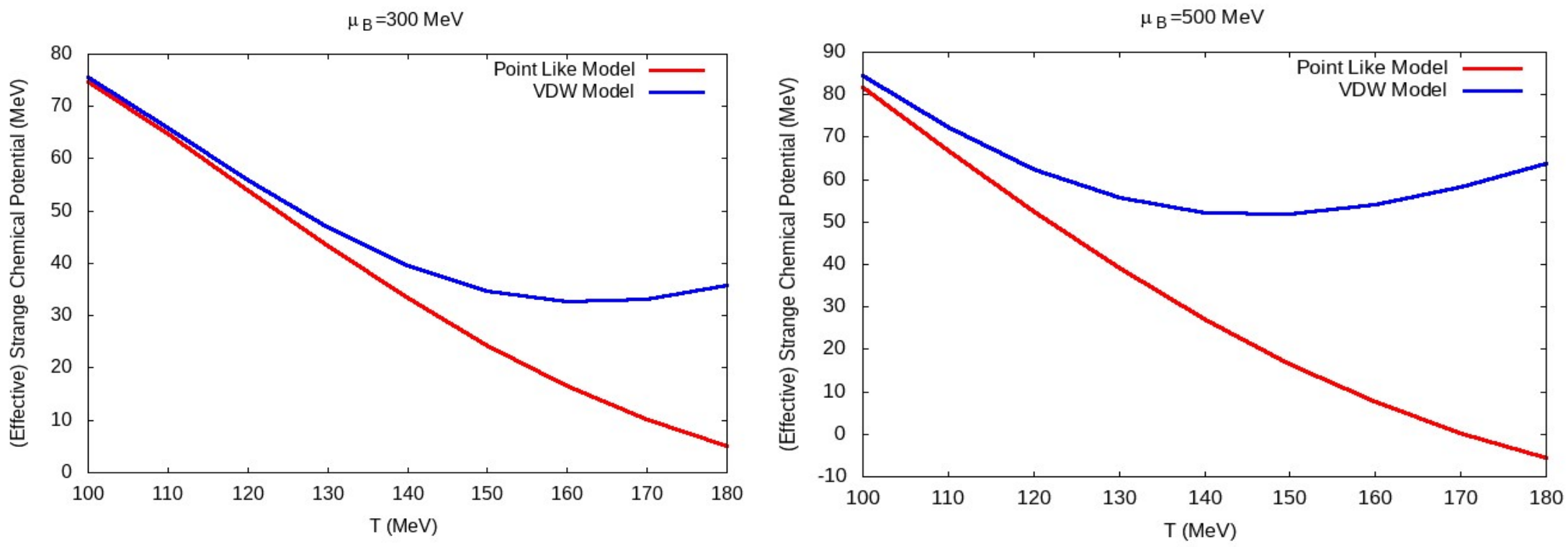

**Fig 4: Variation of "effective" strange chemical potential $(\mu_S^*)$ with temperature at fixed baryon chemical potentials $\mu_B = 300$ MeV and 500 MeV.**

The antistrange to strange hyperon ratios using NIHRG as well as IHRG equation of states have been calculated. This will also enable us to understand that to what extent the "effective" baryon as well as the corresponding "effective" strange chemical potentials i.e. $\mu_B^*$ and $\mu_S^*$ can affect these ratios.

With this purpose we have first calculated the singly antistrange to strange particle ratio viz. $\left(\frac{\bar{\Lambda}}{\Lambda}\right)$.

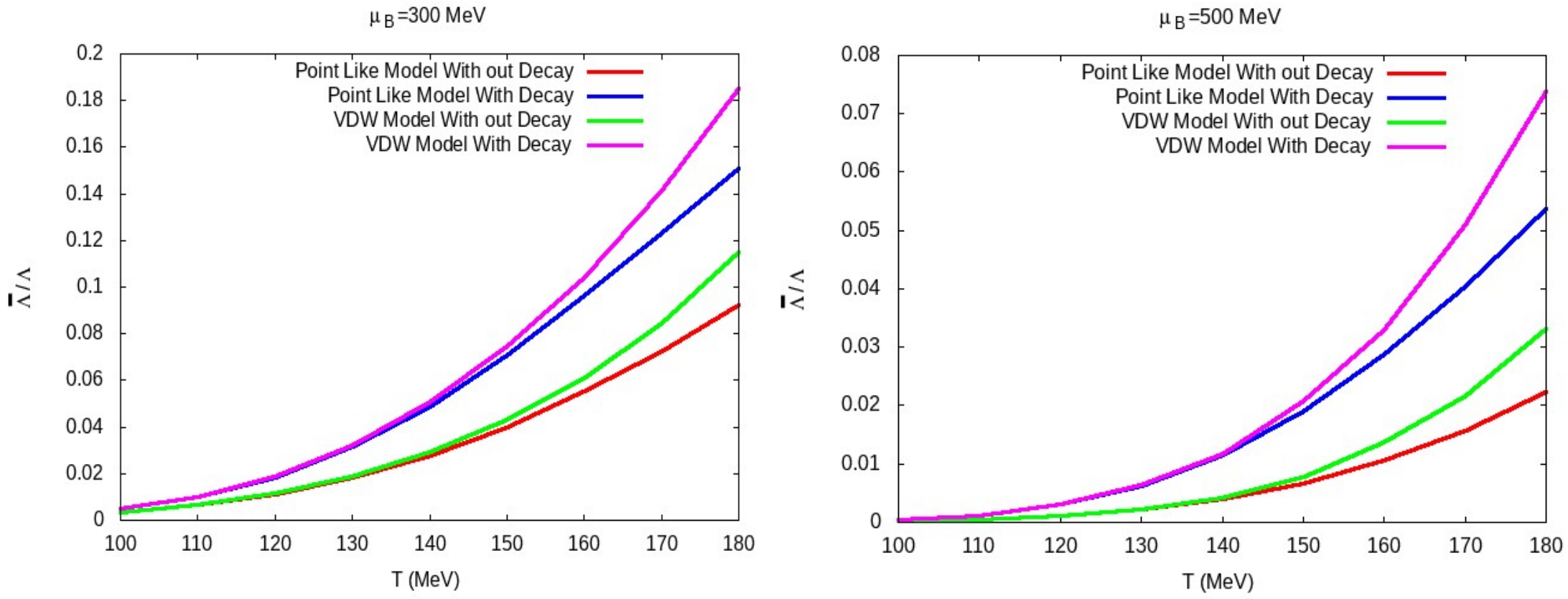

**Fig 5: Dependence of the ratio $\left(\frac{\bar{\Lambda}}{\Lambda}\right)$ on temperature at fixed baryon chemical potential $\mu_B =$ 300 MeV and 500 MeV.**

The curves in figure 5 show that inclusion of the resonance decay contribution can play a very important role here. A significant increase in the $\left(\frac{\bar{\Lambda}}{\Lambda}\right)$ ratio is observed with decay contributions taken into account for both cases i.e. IHRG and NIHRG. For the NIHRG case the ratio shows significant increase for temperatures above 160 MeV. The available experimental data at mid-rapidity from NA49 collaboration [80] gives the values of this ratio at 40A GeV and 80A GeV in CERN-SPS experiments as $\sim$ 0.027 and 0.078, respectively. These can be well reproduced by using NIHRG model using $\mu_B =$ 500 MeV and 300 MeV, respectively for temperatures close to $\sim$ 155 - 160 MeV.

Measurement of "relative" strange hadron yield provides a better understanding of the strangeness production mechanism in nucleus-nucleus collisions. Using the IHRG as well as the NIHRG models we have calculated the $\frac{\Lambda}{p}$ ratio and in figure 6 shown its dependence on the temperature of the system for $\mu_B =$ 300 MeV and 500 MeV.

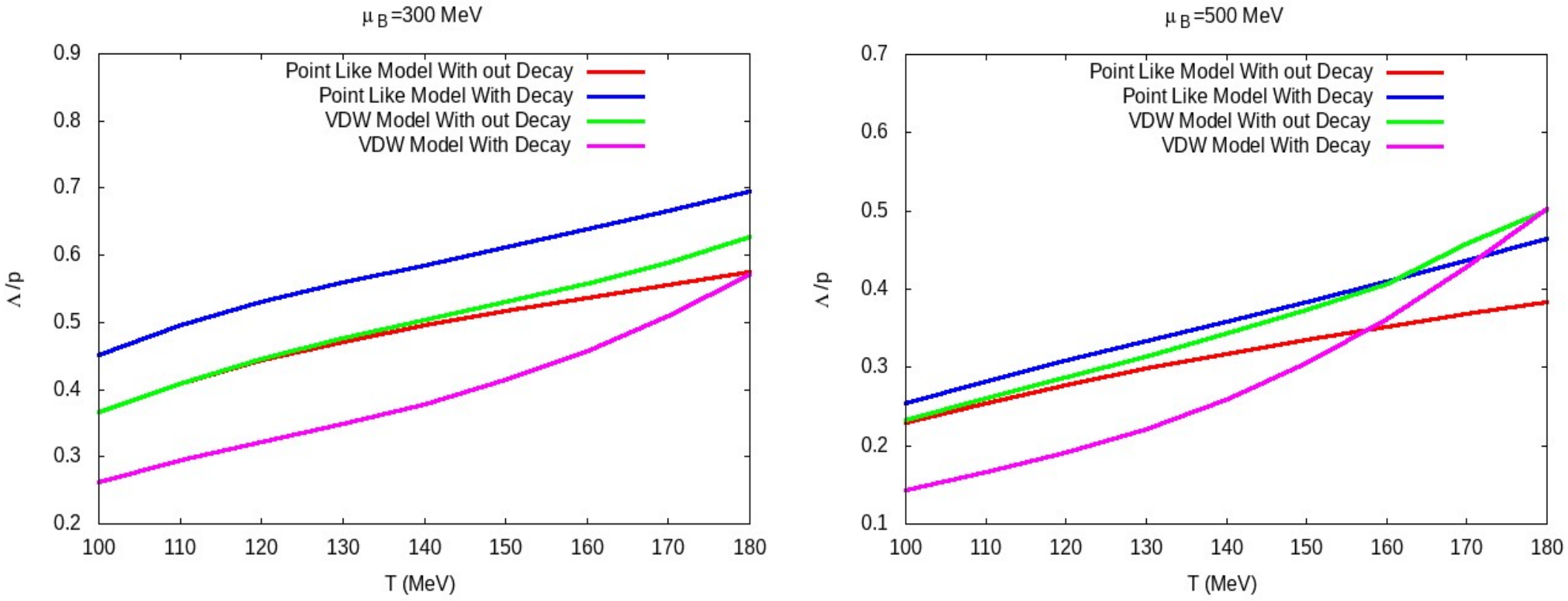

**Fig 6: Dependence of the ratio $\left(\frac{\Lambda}{p}\right)$ on freeze-out temperature at fixed baryon chemical potential $\mu_B = 300$ MeV and 500 MeV.**

It is interesting to see that using the EoS of IHRG the inclusion of resonance decay contribution leads to an enhancement of the $\frac{\Lambda}{p}$ ratio while the behaviour of the system using EoS of the NIHRG the opposite is observed. The values of this ratio at 40A GeV and 80A GeV in CERN-SPS experiments [69, 80] turn out to be about 0.37 and 0.44, respectively. In the present calculation these can be reproduced by the NIHRG model results quite well for $\mu_B = 500$ MeV and 300 MeV, respectively for temperatures close to $\sim$ 165 MeV taking into account the decay contributions. The IHRG model with contribution of the decay products taken into account is seen to over predict this ratio and can reproduce the experimental values only at much lower temperatures.

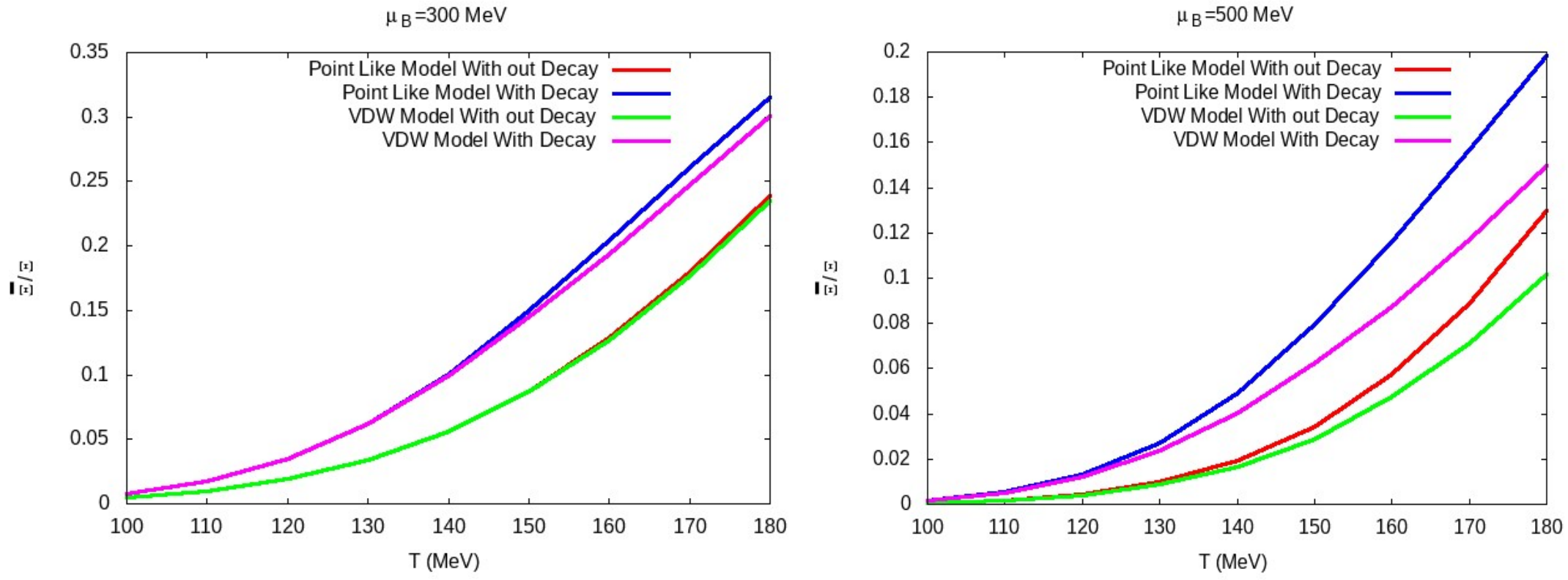

**Fig 7: Ratio of doubly (anti)strange paricles, $\left(\frac{\bar{\Xi}}{\Xi}\right)$ dependence on temperature for baryon rich systems.**

Figure 7 shows the variation of the doubly (anti)strange particle ratios, i.e. anticascade to cascade, $\left(\frac{\bar{\Xi}}{\Xi}\right)$. In this case we notice that when we use Van der Waals (VDW) type EoS there is a slight decrease in the $\left(\frac{\bar{\Xi}}{\Xi}\right)$ ratio compared to the point like hadron case at high temperatures, especially for a baryon rich system. The available experimental data at 40A GeV in Pb + Pb collision gives this ratio ~ 0.07 [81] which can be well described around T ~ 155 MeV in case of NIHRG and ~ 145 MeV for IHRG.

In almost all the above cases we find that the NIHRG EoS can consistently reproduce the experimental particle ratios for the two CERN-SPS energies in the temperature range of 155 – 165 MeV which appears quite reasonable.

Another interesting particle ratio viz. $\left(\frac{K^-}{K^+}\right)$ has been shown in figure 8. We find that for both values of chemical potential the ratio for the two cases i.e. IHRG and NIHRG, initially decreases. But for the case of NIHRG it shows a rising trend after certain value of temperature. This behaviour of NIHRG with VDW type interaction is in contrast with that of the IHRG. Further, the $\frac{K^-}{K^+}$ ratio is seen to be suppressed when heavier hadronic resonance decay contributions are taken into account, especially at smaller temperatures. But at higher temperatures the NIHRG result with decay contributions taken into account leads all the other three cases.

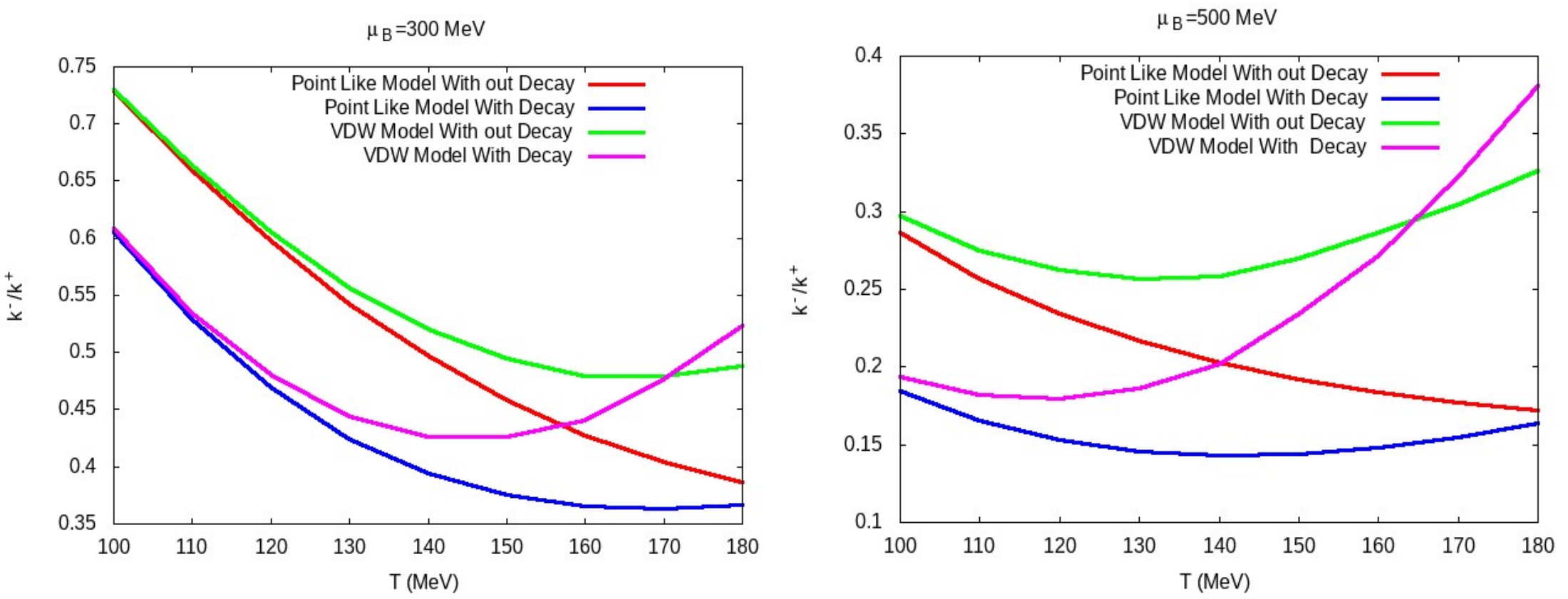

**Fig 8: The $\left(\frac{K^-}{K^+}\right)$ ratio with temperature for baryon rich system using $\mu_B = 300$ MeV and 500 MeV**

The available data from NA49 at 80A GeV from the Pb + Pb collision [82] gives ≈ 0.37, which is seen reasonably close to the IHRG curve for $\mu_B = 300$ MeV when T ~ 145 MeV provided resonance decay contributions are taken into account. Without these contributions the IHRG result highly over predicts the experimental data. At this value of $\mu_B$ the NIHRG result with resonance contributions taken is also seen to somewhat over predict this value. In NIHRG case however one can obtain this value of the $\left(\frac{K^-}{K^+}\right)$ ratio at nearly the same temperatures ($\sim$ 155 – 165 MeV) used for the other particle ratios above if we use a somewhat larger value of $\mu_B \sim$ 350 - 375 MeV.

## 4. Summary and Conclusion:

Within the framework of a statistical thermal model we have used a grand canonical ensemble formulation for a multi-component non ideal hadron resonance gas model. We have considered the attractive as well as repulsive interaction among the constituent baryons (antibaryons). The particle number densities are obtained in a thermodynamically consistent manner. Using the present formulation we have calculated several particle ratios like $\frac{\bar{p}}{p}$, $\frac{\bar{\Lambda}}{\Lambda}$, $\frac{\Lambda}{p}$, $\frac{\bar{\Xi}}{\Xi}$ and $\frac{K^-}{K^+}$. The dependence of these relative particle yields, the "effective" baryon chemical potential ($\mu_B^*$) and the "effective" strange chemical potential ($\mu_S^*$) on T and $\mu_B$ has been studied. We find that the particle ratios get modified particularly at higher values of T and $\mu_B$ by using Van der Waals type EoS i.e. considering the existence of a non ideal hadronic resonance gas (NIHRG) at the hadronic fireball freeze-out which is assumed to be in a state of reasonably high degree of thermo-chemical equilibrium. The application of the NIHRG equation of state shows that these interactions can play important role in describing the relative particle yields especially for a hot baryon rich system. We find that by considering the formation of a NIHRG system in the ultra-relativistic nucleus-nucleus collisions we can quite reasonably predict several experimental particle ratios obtained in the CERN SPS at 80A and 40A GeV within a temperature range of 155 – 165 MeV and choosing $\mu_B$ values 300 MeV and 500 MeV for the two cases, respectively. Elaborate analyses of the hadronic yields for a baryon rich system can be carried out further when data from the upcoming Compressed Baryonic Matter from the FAIR experiments will become available.